\documentstyle[iopconf1]{article}
\begin{document}
\title{Preon Trinity - a new model of leptons and quarks
\footnote{Presented
by S.F. at Beyond 99, Tegernsee, Germany, June 7-11, 1999;
to be published in the Proceedings}
}

\author{Jean-Jacques Dugne\dag, 
\underline{Sverker Fredriksson}\ddag,
Johan Hansson\ddag \ and Enrico Predazzi\S}

\affil{\dag\ Laboratoire de Physique Corpusculaire,
Universit\'{e} Blaise Pascal, Clermont-Ferrand II,
FR-63177  Aubi\`{e}re, France}

\affil{\ddag\ Department of Physics, Lule\aa \ University
of Technology,\\
SE-97187 Lule\aa , Sweden}

\affil{\S\ Department of Theoretical Physics,
Universit\'{a} di Torino,\\
IT-10125 Torino, Italy}

\beginabstract
A new model
for the substructure of
quarks, leptons and weak gauge bosons
is discussed. It is based on three fundamental
and absolutely stable
spin-$1/2$ preons. Its preon flavour $SU(3)$
symmetry leads to
a prediction of nine quarks, nine
leptons and nine heavy vector bosons.
One of the quarks has charge $-4e/3$,
and is speculated to be the top quark
(whose charge has not been measured).
The flavour symmetry leads to
three conserved lepton numbers in all
known weak processes, except for some
neutrinos, which might
either oscillate or decay.
There is also a (Cabibbo) mixing of the
$d$ and $s$ quarks due to an internal
preon-antipreon annihilation channel.
An identical channel exists inside
the composite $Z^0$, leading to
a relation between the Cabibbo and
Weinberg mixing angles.

\endabstract

\section{Introduction}

In this talk I will present a new preon model
where leptons, quarks and heavy vector bosons
are composite objects with a fairly simple
inner structure. It is the
result of a collaboration between the
universities of Clermont-Ferrand, Torino and
Lule\aa \ within an EU-sponsored network,
where also groups from Paris and Thessaloniki
are involved in other projects.

Our inspiration has come from a certain
fatigue with the standard model
and its rather complicated Higgs mechanism,
but also from its phenomenological
success, which has convinced us that
``there is something more fundamental behind''.
Let me therefore start by giving our
favourite preon arguments, and first two
of a more general nature:

\vspace{2mm}

{\it Why not?}\\
It should be fully natural to speculate
about compositeness, especially as it
is impossible to prove, by any means,
that a particle is {\it not} composite!
So, why not composite preons?
Well, pre-preons already exist
(\ldots in the literature).

\vspace{2mm}

{\it There is always a deeper layer\ldots}\\
History tells us that whenever
a model of matter has become more and more complicated
without providing a deeper understanding,
another layer of fundamental particles has been
suggested - and found. But another
lesson of history is that a vast majority
of all researchers have treated the currently
known fundamental particles as {\it the} ultimate
level, until the facts have been obvious!

\vspace{2mm}

Among the more detailed arguments, we would
like to list the following (many of which,
but not all, have
been discussed before \cite{preons}):

(i) There are by now ``too many''
leptons and quarks.
It is a puzzle why God the Almighty
would prefer twelve fundamental particles.
The only reason I can find is that he
chose one per Apostle, but that would
restrict the standard model to Christianity,
which seems a too limited scope. {\it Three} is
a more universally sacred number.

(ii) There is a pattern among
the six leptons and six quarks, {\it i.e.},
the family structure. Historically,
such patterns
have led to ideas about compositeness, with the
quark model as a typical example.

(iii) The mathematically
least attractive features
of the standard model come
from the fact that the weak interaction
has gauge bosons that are massive (as well
as unstable and of different charges).
In preon models, they can
be seen as composite states. The weak interaction is
a ``leakage'' of  multi-preon
states, in analogy to the
nuclear force being mediated by
quark-antiquark states. If so, there is no
fundamental electroweak unification and
no need for the Higgs mechanism (nor for multi-Higgs,
higgsinos or composite Higgs).

(iv) An overlooked hint
is that most quarks and
leptons are {\it unstable},
which in our view disqualifies them as fundamental
particles. Historically, all decays of
``elementary'' particles have sooner
or later been interpreted in terms of
more fundamental objects.

(v) There are mixings, transitions,
or oscillations
among some of the ``fundamental'' particles,
such as among some quarks, and also
among neutrinos. This reminds about
the mixing of $K_{0}$ and $\bar{K}_{0}$,
now being understood in terms of
quarks.

(vi) There are conserved quantum numbers
such as lepton numbers and weak isospin,
which we do not understand. When isospin and
strangeness were introduced in particle
physics, the explanation turned out to be
in terms of quarks and (partly)
conserved quark flavours.
One of the visions for any preon model
should be to reduce the flora of
{\it ad hoc} quantum numbers.

\section{General preon considerations}

Preon models have so far focused on explaining
the lightest family of two quarks ($u$ and $d$)
and two leptons ($e$ and $\nu_e$)
with as few preons as possible. This means either
a minimal number of (two) different
preons, {\it e.g.}, the ``rishons''
\cite{harari,shupe}, or a minimal number
of (two) preons inside a quark or lepton,
{\it e.g.}, the ``haplons'' \cite{fritzsch}.

The two heavier families are either
considered ``excitations''
of the light one, or prescribed to
have a completely different internal
structure, {\it e.g.}, with different
numbers of preons inside different
quarks. As a result, no simple, consistent
and predictive preon model exists. 
Neither have preon models been
able to explain the cornerstones of the
standard model from preon
properties. Rather, lepton number
conservation, the Cabibbo-Kobayashi-Maskawa (CKM)
mixings \cite{CKM}, and the similarities
between leptons and quarks,
have either been left unexplained, or
implemented on the preon level.

Nevertheless, we have been much inspired
by the work of others, when formulating
our model. In particular, we would like
to mention that we have stolen, and
customised, the following ideas:

(i) the idea from the original $SU(3)$ quark model
by Gell-Mann and Zweig \cite{quarks} that
all hadrons known in the early 1960s can be
explained in terms of only three light quarks,
with a (broken) flavour-$SU(3)$ symmetry;

(ii) the idea from the rishon model
that members of the lightest
family have three preons each;

(iii) the idea from the haplon model
that the members of the lightest family can
be built by one spin-$1/2$ and one spin-$0$
object;

(iv) the idea of diquark models \cite{anselm}
that quarks like to pair-up in tightly bound
spin-$0$ systems;

(v) the idea of supersymmetry that spin-$1/2$
objects have relatives with spin $0$, even if only
in a phenomenological sense, such as the
``supersymmetry'' between quarks and diquarks
in mesons and baryons \cite{anselm}.

\section{A trinity of preons}

A preon model for six leptons
and six quarks must have at least
five different preons in the sense of
the haplon model, {\it i.e.}, three
with spin $1/2$ and two with spin $0$,
or {\it vice versa}.
In order to get a more symmetric scheme
it seems reasonable to have three of each,
giving nine leptons
and nine quarks.

The spin-$1/2$ and spin-$0$
preons can be given pairwise identical charges,
which by ``accident'' are those of the three
lightest quarks. In order to bring down the
number of preons, we suggest that the spin-$0$
ones are not fundamental, but tightly bound
``dipreon'' pairs of the spin-$1/2$ preons.
Calling the preons $\alpha$, $\beta$ and $\delta$,
and giving them charges $+e/3$, $-2e/3$ and
$+e/3$ (for simplicity; there is an ambiguity
between the names preon and antipreon), we
get the simple and symmetric
scheme in Table 1. Each preon has a partner,
which is the dipreon formed by the other two
(anti)preons.

\begin{table}
\begin{center}
\caption{Our ``supersymmetric'' preon scheme.}
\begin{tabular}{l|ccc}
\topline
charge & $+e/3$ & $-2e/3$ & $+e/3$\\
\midline
spin-$1/2$ preons & $\alpha $ & $\beta $ & $\delta $ \\
spin-$0$ (anti)dipreons & $(\bar{\beta} \bar{\delta})$ &
$(\bar{\alpha} \bar{\delta})$ &
$(\bar{\alpha} \bar{\beta})$\\
\bottomline
\end{tabular}
\end{center}
\end{table}

This model leads to a
surprisingly rich spectrum of
predictions and explanations connected
to the standard model. But
before building up leptons, quarks and
vector bosons by combining preons with
dipreons and/or their anti-particles,
we list the properties of the preons
themselves.

(i) {\it Mass}:
In order to understand why only
six leptons, six quarks and three
vector bosons have been seen, it is
tempting to speculate that one of
the preons is superheavy, say the $\delta$.
The $\alpha$ and $\beta$ are
much lighter.
The ``heaviness'' of the $\delta$ preon
hence plays the role of strangeness
in the quark model.
It should be noted though, that the more
bound a fundamental particle is, the less
precise is the definition of its mass.
An illustrative paradox
is that dipreons containing the $\delta$
do not seem to be superheavy.

(ii) {\it Spin and electric charge}:
These are just implemented on the preon level.
It might be worthwhile here to contemplate
the fundamental difference between the
$\alpha$ and $\delta$ preons. Maybe this is
in terms of a different magnetic charge,
coupled to both electric charge and spin.

(iii) {\it QCD colour}:
It is necessary to assume that all preons
are colour-${\bf 3^*}$ in normal QCD, in order
to understand why leptons and vector bosons
are colour
singlets and quarks colour triplets.
The assignment ${\bf 3^*}$ instead
of ${\bf 3}$ is again a technicality because
of our definition of preons {\it vs.} antipreons.
Then also the (anti)dipreons in
Table 1 are ${\bf 3^*}$
($\bf 3 \otimes \bf 3 = \bf 3^* \oplus \bf 6$).

(iv) {\it Preon flavour}:
The preon flavour plays, at first sight,
the same role in this model as the
quark flavours did in the original
quark model, {\it i.e.}, the model
has an approximate flavour-$SU(3)$ symmetry.
It cannot be exact for masses and
wave functions, but unlike the quark model,
we suggest that the net flavours are
exactly conserved in nature, as a
consequence of the next point.
Note that the preons are flavour-$\bf 3$,
the dipreons $\bf 3^*$, and
the antidipreons $\bf 3$.

(v) {\it Absolute stability!}:
Unlike the situation for ``fundamental''
particles in the conventional standard
model, we assume that preons are absolutely
stable against decay. All weak decays in
nature are therefore consequences of a mere
reshuffling of preons into particles with
a lower total rest mass.

(vi) {\it ``Hypercolour''?}:
Most preon models rely on some new,
superstrong force that keeps leptons
and quarks together. Assuming that it
is QCD like, it is usually called
hyper-QCD or super-QCD, with hyper-gluons
etc. There are also suggestions that it
might not be $SU(3)$ symmetric, but obey
some more complicated group theory, such
as $SU(4)$.

It is noteworthy that {\it no} other preon
quantum properties are needed in order to
understand all quantum numbers of leptons and quark.
So, there is no lepton number, no baryon number,
no isospin, strangeness or weak isospin on
the preon level.

\section{The leptons}

Leptons are assumed to be three-preon
states, in the form of a preon and
a dipreon, all in colour-singlet
($\bf 3^{*} \otimes \bf 3 = \bf 1 \oplus \bf 8$).
We assume that colour-octet leptons
do not exist.
The lepton scheme that we favour is given
in Table 2.

Note that this matrix is set up as a rough scheme,
without a deeper consideration of what the
actual preon wave functions would be, {\it i.e.},
quantum-mechanical mixings of the simple products
of single-preon wave functions. We will turn back
to such questions later. Next we discuss the
lepton properties, point by point.

\begin{table}
\begin{center}
\caption{Leptons as three-preon states.}
\begin{tabular}{c|ccc}
\topline
& $(\beta \delta)$
& $(\alpha \delta)$
& $(\alpha \beta)$ \\
\midline
$\alpha$
& $\nu_{e}$
& $\mu^+$ & $\nu_{\tau}$ \\
$\beta$ & $e^-$
& $\bar{\nu}_{\mu}$
& $\tau^-$ \\
$\delta$
& $\bar{\nu}_{\kappa 1}$
& $\kappa^+$
& $\bar{\nu}_{\kappa 2}$ \\
\bottomline
\end{tabular}
\end{center}
\end{table}

(i) All leptons are correctly reproduced
with all their quantum numbers (except,
possibly, helicity,
since it is not obvious from the scheme why
neutrinos are left-handed and antineutrinos
right-handed).
There is an ambiguity in the scheme,
since it would work equally well with the
``electron'' column interchanged with the
``$\tau$'' column. We will discuss this
ambiguity later.

(ii) There are three new (superheavy)
leptons, all containing an ``isolated''
$\delta$ preon. Two of these are neutrinos,
which must be heavier than half the
$Z^{0}$ mass, in order not to violate
the finding that there are only three light
neutrinos. These neutrinos must naturally
be unstable.

(iii) There is no subscheme
in terms of three lepton families.
The only mathematical
structure is the one given by preon-flavour
$SU(3)$, which splits up the nonet in one
octet and one singlet, exactly like with
the three-quark baryon octet
($\bf 3 \otimes 3^* = 8 \oplus 1$).
When constructing true
flavour-$SU(3)$ based wave functions
the singlet is to be found as a linear
combination of the three neutrino cells
on the main diagonal, while the other
two combinations
are the lepton equivalents of the baryons
$\Sigma^{0}$ and $\Lambda^{0}$ in the
quark model.

(iv) The basic scheme contains the
electron and the {\it anti}muon on equal
footing, and similarly for their neutrinos.
This whets the appetite for speculations that
helicity has to do with the properties of the
``naked'' preon outside the spin-$0$ dipreon.

(v) There is no simple mass-ordering in the
scheme, which makes it a much tougher challenge
to find mass-formulas and a dynamics, than in
the quark model. One trend is that
the mass increases along all diagonal lines from upper
left to lower right.

(vi) The three traditional lepton numbers are
conserved in all well-established
processes. This mirrors the conservation of
preon flavour. As to the
best of our knowledge, this is the only
preon model that
links the lepton numbers, and the virtual
three-family structure, to a more
fundamental concept.
Muon decay, $\mu^- \rightarrow
\nu_{\mu} + e^- + \bar{\nu}_e$,
is a typical example:
$\bar{\alpha}
(\bar{\alpha} \bar{\delta}) \rightarrow
\bar{\beta} (\bar{\alpha} \bar{\delta}) +
\beta (\beta \delta) +
\bar{\alpha} (\bar{\beta} \bar{\delta})$.
The dipreon goes into
the muon neutrino, while the preon hides
inside a $W^- = \beta \bar{\alpha}$,
which then decays into the electron
and its neutrino.
All efforts to violate lepton number
conservation
by splitting up the leptons differently
lead to final states that
violate energy conservation.

(vii) There is no general lepton number conservation!
First of all, there is no fourth lepton number
connected to the superheavy $\kappa$ and its two
neutrinos. Rather, their decays {\it must} violate
also the normal lepton numbers,
an illustrative example being:
$\kappa^+ \rightarrow \mu^{+} +
\nu_e + \bar{\nu}_{\tau}$.

(viii) An attractive aspect
of the lack
of a general lepton number conservation
is that the three neutrinos, $\nu_e$,
$\bar{\nu}_{\mu}$ and
$\bar{\nu}_{\kappa 2}$ along the
main diagonal can mix quantum-mechanically,
because they have
identical preon contents, differing only
in the internal spin composition:
$\nu_e = \alpha (\beta \delta)$,
$\bar{\nu}_{\mu} = \beta (\alpha \delta)$
and $\bar{\nu}_{\kappa 2} =
\delta (\alpha \beta)$.
This brings up the issue of the
ambiguity between the left-most and
right-most columns in Table 2, since a shift
of columns would involve the $\nu_{\tau}$
in the mixing, instead of the $\nu_{e}$.
A three-neutrino mixing might induce
neutrino oscillations, $\nu_e \leftrightarrow
\bar{\nu}_{\mu} \leftrightarrow \bar{\nu}_{\kappa 2}$.
But there could also be decays,
like the electromagnetic
$\bar{\nu}_{\mu} \rightarrow
\nu_e + \gamma$, or the more exotic
$\bar{\nu}_{\mu} \rightarrow
\nu_e + \phi$, where $\phi$ is a light
scalar, such as a Goldstone boson, or
a bound scalar preon-antipreon
state, say $(\alpha \bar{\alpha})$.
The photon decay has been studied recently
by one of us within the same formalism as
the quark analogy
$\Sigma^0 \rightarrow \Lambda^0 + \gamma$
\cite{hansson}, but no conclusion can be drawn
about the neutrino lifetime or the photon
spectrum.
Finally, it should be noted that it is the
{\it anti}muon neutrino that mixes with the
electron (or tau) neutrino.
This is not obviously
forbidden by helicity conservation, since
massive neutrinos have both left-handed
and right-handed components. One can also
think of helicity-conserving oscillations,
where the neutrino appearing with the wrong
helicity is sterile. Such sterile neutrinos
are often needed in ambitious efforts to fit
all neutrino data in terms of oscillations.

(ix) A final, but important point, is that
the lack of lepton number conservation cannot,
in our scheme, allow for mixings of the {\it charged}
leptons. Such mixings can occur {\it only} if there is
a more basic, quantum-mechanical
(``Cabibbo'') mixing
of the $\alpha$ and $\delta$ preons inside leptons
and quarks. That would violate the exact
conservation of three preon flavours,
but cannot be excluded, on the {\it per mille} level,
by the experimental data. We will discuss this issue
again in connection to quarks.

\section{The quarks}

In a similar spirit as with the leptons,
we now construct, in Table 3, the quarks as
bound states of a preon and an antidipreon
in colour triplet
($\bf 3^* \otimes 3^* = 3 \oplus 6^*$).
Again, the contents of
the cells are not meant to give the
exact preon wave functions of the quarks.

\begin{table}
\begin{center}
\caption{Quarks as preon-antidipreon states.}
\begin{tabular}{l|ccc}
\topline
& $(\bar{\beta} \bar{\delta})$
& $(\bar{\alpha} \bar{\delta})$
& $(\bar{\alpha} \bar{\beta})$  \\
\midline
$\alpha$
& $u$
& $s$ & $c$ \\
$\beta$
& $d$
& $X$ & $b$  \\
$\delta$
& $t?$
& $g$ & $h$  \\
\bottomline
\end{tabular}
\end{center}
\end{table}

The following list of quark properties takes up
several new features that do not appear for leptons.

(i) All known quarks are correctly reproduced.
There is no overall ambiguity in the scheme.

(ii) There are three new
quarks, but these are not obviously
the ones with an ``isolated'', superheavy
$\delta$ preon. Two of them belong to
this category, the $g$ and $h$ quarks
(for ``gross'' and ``heavy''), but the third
($X$) is on the second line, and not clearly
superheavy.

(iii) On the other hand, the top quark
is among the superheavies,
which would explain why it is so much
heavier than its conventionally
assigned partners,
the $b$ quark and the $\tau$ lepton.
This would also give some hope
that ``superheavy'' in our model
means around a few hundred GeV, which
would make the CERN LHC ideal
for discovering the remaining
leptons, quarks and vector bosons.
They might even be within reach
of the Fermilab Tevatron.

(iv) The quarks have no three-family
grouping either, but here the flavour-$SU(3)$
decomposition is into a sextet and an
anti-triplet
($\bf 3 \otimes \bf 3 = \bf 6 \oplus \bf 3^*$).
The anti-triplet contains the three
quarks on the main diagonal, while the
other quarks form the sextet.
Here it should be mentioned that
a flavour-sextet assignment for
quarks has been suggested also by
Davidson {\it et al.} \cite{davidson},
and speculated to be a consequence
of a preon substructure, although
without reference to any particular
preon model.

(v) The appearance of the
$X$ quark with charge $-4e/3$ opens
up for some thrilling speculations.
At first sight it should not be
superheavy, since it does not
have a naked $\delta$. However,
the $X$ belongs
to a flavour antitriplet together with
the $u$ and $h$, and different
$SU(3)$ representations can have
different mass formulas.
Another idea is that a ``light'' $X$ quark
has escaped discovery. Searches for
new quarks seem to focus on
a ``fourth-family'' $b'$ with
charge $-e/3$ \cite{rpp}.
They rely on model-dependent
assumptions on $b'$ decay \cite{abachi}
in ways that would make the suggested
decay of the $X$ ``invisible''.
Neither has there been searches
for new resonances in $e^+e^-$ collisions
at high energies in the traditional way of
fine-tuning the total energy in small steps.
It is, however, unlikely
that a light $X$ would not have been detected
in the search for the top quark \cite{top}.
The case for an $X$ with a mass below,
say, the $W$ mass is therefore weak.

(vi) A final speculation about the
$X$ is that it is indeed identical
to the discovered top quark.
This idea cannot easily be dismissed, since there
is no measurement of the top charge
\cite{top}. The top quark was found
through its presumed main decay channels,
namely semi-leptonic ones like
$t \rightarrow b + \ell^+ + \nu$, or
non-leptonic ones like
$t \rightarrow b + u + \bar{d}$,
where a few $b$ quarks have
been ``tagged'' by a charged muon
in semi-leptonic decays.
However, the situation is rather
complex because a total event contains
the decay of the full $t \bar{t}$ pair
into several leptons and hadronic jets,
and the identification and matching of
those are non-trivial. The corresponding decay
channels for the antiquark $\bar{X}$ would be
$\bar{X} \rightarrow \bar{b} + \ell^+ + \nu$
and $\bar{X} \rightarrow \bar{b} + u + \bar{d}$,
so that the full $X \bar{X}$ decay would
give the same leptons and jets
as in a $t \bar{t}$ decay, although with
a different $b-W$ matching.
A study along these lines \cite{bellettini} shows
that only a couple of events give a weak
support for the conventional charge,
in the sense of
$\chi^2$ fits to different jet identifications,
while the bulk of events are inconclusive.
Hopefully, this issue will be settled soon
at the upgraded Tevatron, thanks to better
statistics and also better possibilities to study
outgoing particles much closer to the
interaction vertex.
It is interesting to note that also Chang {\it et al.}
\cite{chang} have suggested that the top charge
is $-4e/3$, although within an analysis
built on details of the standard model that do
not appear in our preon model ({\it e.g.}, the
Higgs mechanism).

(vii) Quark decays are mostly similar to
lepton decays, since they normally mean
a regrouping of preons. An example is beta decay:
$d \rightarrow u + e^- + \bar{\nu}_{e}$,
which is
$\beta(\bar{\beta}\bar{\delta}) \rightarrow
\alpha(\bar{\beta}\bar{\delta}) +
\beta(\beta \delta) +
\bar{\alpha}(\bar{\beta}\bar{\delta})$.
However, some quarks have pairwise identical
net preon flavour, just like the three
neutrinos discussed earlier. The best
example is the $d$ and $s$ quarks, having
the flavour of the $\delta$.
This makes possible a quantum-mechanical
mixing of the two quarks, and another
type of transition between them, compared
to normal quark decays.
This is the preonic explanation of
the Cabibbo mechanism. An $s \leftrightarrow d$
transition goes through annihilation channels
like $\alpha \bar{\alpha} \leftrightarrow \beta \bar{\beta}$
involving the ``naked'' preon and the corresponding
antipreon inside the dipreon. No such channels
exist for leptons, but as we will see below,
they are important also inside the preon-antipreon
$Z^0$. In terms of wave functions and the
notions of the Cabibbo theory, the $d$ and $s$
quarks are the flavour-$SU(3)$ eigenstates
produced in preon processes, while the mass
eigenstates, $d'$ and $s'$,
relevant for weak decays, are
mixtures of these preon states,
parametrised by the Cabibbo angle.
As the Cabibbo mechanism is just put in
``by hand'' in the standard model, we regard this
connection between the preon substructure and
the mixing of $d$ and $s$ one of the most
promising features of our model.

(viii) The smaller elements of
the CKM matrix cannot be accounted for by
such preon annihilation channels, although
it is noteworthy that they are much smaller
than the Cabibbo mixing, and hence might
have a different explanation.
We point out
that such effects can result from a
very small quantum-mechanical mixing
of the $\alpha$ and $\delta$ preons
inside the quark (and lepton) wave
functions. This would mean that
flavour-$SU(3)$ is heavily broken
on the mass level (quark and lepton
masses), somewhat broken
in wave functions (Cabibbo mixing), and
just a little broken on the quantum-number
level ({\it i.e.}, preon flavour is not exactly
conserved). Again, the situation is
principally very similar to the early
quark model.

\section{The heavy vector bosons}

The vector bosons are preon-antipreon
states, as in several other preon models.
In our model there will be nine of them,
as in Table 4.

\begin{table}
\begin{center}
\caption{Vector bosons as preon-antipreon states.}
\begin{tabular}{l|ccc}
\topline
& $\bar{\alpha}$
& $\bar{\beta}$
& $\bar{\delta}$  \\
\midline
$\alpha$
& $Z,Z'$ & $W^+$ & $Z'''$ \\
$\beta$
& $W^-$ & $Z',Z$ & $W'^-$  \\
$\delta$
& $Z'''$ & $W'^+$ & $Z''$  \\
\bottomline
\end{tabular}
\end{center}
\end{table}

The following observations can be made.

(i) This nonet is very similar to the
vector mesons in the quark model
($\rho,\omega,\phi,K^*$). Some examples are:
Both carry a ``leakage force'',
the weak and nuclear ones;
both $Z^0$ and $\rho^0$ are
heavier than their charged counterparts
(which is rare in particle physics);
and both $Z^0$ and $\rho^0$
are believed to mix with the photon,
as described by the electroweak theory 
and the vector meson dominance model.
In the meson case it is acknowledged that
these phenomena are consequences of the
quark structure, while for vector bosons
they are just put in by hand in the standard model,
but are due to preons in our model.

(ii) There are no known scalar counterparts to the
vector bosons, although these are expected in
many preon models \cite{preons}. It is not known
if they are even heavier than the vector bosons, or have
extremely weak couplings to other particles,
or are very light, but hidden inside scalar
mesons as Fock states. It is a crucial challenge to
preon models to find these scalars.

(iii) The standard model concepts of weak isospin,
Weinberg mixing etc appear in this model on the same
level as the Cabibbo and neutrino mixings, {\it i.e.},
as consequences of (broken) flavour-$SU(3)$
on the preon level. In Table 4, the hypothetical particle
normally called $W^0$ (the weak isotriplet) is the
$SU(3)$ eigenstate $\alpha \bar{\alpha} -
\beta \bar{\beta}$, while the isosinglet $B^0$
is $\alpha \bar{\alpha} +
\beta \bar{\beta}$ (neglecting a possible
admixture of the superheavy $\delta$).
The mass eigenstates, $Z$ and $Z'$, are mixtures
of these, parametrised by the Weinberg angle.
Note that the state orthogonal to $Z$ is {\it not}
the photon, as in the standard model, but the
heavier $Z'$. This is more like the vector mesons,
where the $\omega$ is the heavier partner of
the $\rho^0$. The
difference between the $W^0$ and $B^0$ components
of $Z$ can be seen through decays, where the latter
mode gives access to preon-antipreon annihilation
channels, identical to the ones inside quarks.
This makes it extremely tempting to speculate
about a possible relation between the Cabibbo
angle for quarks and the Weinberg angle for
$Z$ decays. We have found such a simple relation, but
there are so many ifs and buts in its proof
that we will present
it in a publication of its own.

\section{Further challenges}

Our model provides
a qualitative understanding of
several phenomena that
are not normally analysed within preon models,
nor within the standard model.
It can hence serve as a basis for a
deeper understanding
of the quantitative success of
the standard model.

However, many problems remain to be solved,
and I will finish by listing some of them below.

(i) The model lacks a dynamical basis, which means
that masses cannot yet be explained,
the exact preon wave functions cannot be calculated,
and the forces that keep leptons and quarks together
are not understood. In this respect we are worse off
than the original quark model, which at least
rested on a phenomenologically successful mass formula
for baryons.

(ii) There is no particular reason why quarks would
be as fundamental as leptons in our model, or even
exist as particles, since there is no obvious
hyper-QCD theory that would make quarks ``hyper-colour
neutral'', like leptons. The fundamental
status of quarks must therefore be due to some more
intricate dynamics.

(iii) It is not yet clear to us if {\it all}
phenomenologically successful aspects of
the electroweak theory can be explained by preons,
not to mention CP violation. A lot of work remains to
be done, maybe of the order of the total work so far
devoted to the standard model, {\it i.e.}, a few 
thousand man-years!

(iv) The hints of a slightly broken preon flavour
conservation reminds so much of the quark model that
the mind goes to yet more preons, even unstable ones
and in families,
which would again call for a model with fewer, and
more basic constituents, {\it i.e.}, pre-preons.

\section*{Acknowledgements}

I would like to thank the Organisers of
this meeting, and the audience, for providing a wonderful
intellectual atmosphere, and a breathtaking environment.
We acknowledge helpful and inspiring remarks
by P. Arve and D. Enstr\"om,
illuminating correspondences
with H. Fritzsch, A. Davidson, E. Ma and N. Tracas,
as well as valuable experimental information from
G. Bellettini, R. Partridge and G. VanDalen.
This project is supported by the European
Commission under contract CHRX-CT94-0450,
within the network ``The Fundamental Structure of Matter".

\end{document}